\documentclass[aip,jcp,reprint]{revtex4-1}

\usepackage{graphicx}
\usepackage{amsmath, amsthm}

\begin{document}

\title{Simple approach for calculating the binding free energy of a multivalent particle}
\author{Nicholas B. Tito}
	\affiliation{Department of Chemistry, University of Cambridge, Lensfield Road, Cambridge CB2 1EW, UK}
	\email{nicholas.b.tito@gmail.com}
\author{Stefano Angioletti-Uberti}
	\affiliation{International Research Centre for Soft Matter, Beijing University of Chemical Technology, Beijing, China}
\author{Daan Frenkel}
	\affiliation{Department of Chemistry, University of Cambridge, Lensfield Road, Cambridge CB2 1EW, UK}

\date{\today}

\begin{abstract}
We present a simple yet accurate numerical approach to compute the free energy of binding of multivalent objects on a receptor-coated surface. The method correctly accounts for the fact that one ligand can bind to at most one receptor. The numerical approach is based on a saddle-point approximation to the computation of a complex residue. We compare our theory with the powerful Valence-Limited Interaction Theory (VLIT) (J. Chem. Phys. {\bf 137}, 094108(2012), J. Chem. Phys. {\bf 138}, 021102(2013)) and find excellent agreement in the regime where that theory is expected to work. However, the present approach even works for low receptor/ligand densities, where VLIT breaks down. 
\end{abstract}

\maketitle

\section{Introduction}

Multivalent particles are microscopic entities that can bind to multiple "receptors" via flexible ligands.   For example, a multivalent particle can bind to receptors placed on an membrane (e.g. a cell wall) or a hard surface (e.g. a sensor).\cite{Mammen:1998im, MartinezVeracoechea:2011kn, Michele:2013bw, Varner:2015dh} The particle might be an oligomeric protein complex, a star polymer, a virus, or a functionalised nano-colloid. 

When the particle diffuses over the surface, it samples different local arrangements and concentrations of receptors. The binding free energy of the particle depends on an average over the bound-state partition function in each locality. In general, computing this partition function is non-trivial as it requires knowledge of the configurational entropy of the ligands for all possible local binding arrangements. \cite{Varilly:2012gl}

To simplify matters, we consider a model where the ligands are modelled as non-self-avoiding polymers on a lattice, although the approach will also work off-lattice. We derive a simple yet exact expression for the binding free energy of a multivalent particle, incorporating an analytical average over every possible local receptor configuration. To calculate the free energy in practice, we employ a saddle-point approximation.

In what follows, we consider a multivalent object at height $h$ above a surface, represented in this example as a two-dimensional lattice, although in general the receptors need not be restricted to a plane. 

For a fixed position $h$, the ligands on the object are able to access $N_A$ surface sites. We assume that the receptors are distributed randomly over the $N_A$ surface lattice sites; the probability that $N_R$ of the surface sites are receptors is defined as $P(N_R)$. The distribution $P(N_R)$ can take any form. The receptors are also assumed to be immobile, as binding sites embedded in the surface; however, our approach can be generalised to account for oligomeric or polymeric receptors.

The probability that a ligand binds to a receptor site $j$ depends on the partition function $q_j$, the Boltzmann-weighted sum of all conformations of a ligand that bind to site $j$. We define $f_j$, the free energy of binding to receptor $j$ through $q_j \equiv e^{-\beta f_j}$. Importantly, no two ligands can bind to the same receptor; this condition is enforced strictly in our approach. 

\section{Model}

We now show how to calculate the average binding free energy of the multivalent object at distance $h$ from the surface. We start by considering the situation where $N_R$ receptors are within the range that can be reached by the $N_L$ ligands. Of course, these receptors can be distributed in many different ways over the $N_A$ accessible surface sites. Initially, we will consider only one specific realisation. Subsequently, we will average over all possible receptor distributions.  

Let us first consider the case that $\lambda$ out of the $N_L$ ligands are bound to the receptors.  Clearly, $\lambda\le N_L$ and $\lambda \le N_R$. We can consider the partition function $Q_b$ that accounts for all possible ways to bind the $\lambda$ ligands to the accessible receptors:
\begin{equation}
	Q_b(N_R, \lambda) =  \sum_{\{\lambda\}_{N_R}}{q_j q_{k\neq j} ...q_{\lambda \neq j, k ...}}  \\
	\label{eqn:QbNR}
\end{equation}
The summation is over all possible subsets of $\lambda$ surface sites out of the available $N_R$.

The number of possible ligand-receptor combinations (i.e. the number of terms in Eq. \ref{eqn:QbNR}) can be extremely large even for moderate $N_R$ and $\lambda$. In order to calculate $Q_b$ we employ the residue theorem of complex functions. To begin, we define the auxiliary function
\begin{equation}
	Z(z) = \prod_{j = 1}^{N_R}{\left(1 + z q_j \right)},
	\label{eqn:AuxFunc}
\end{equation}
where $z$ is a complex variable. This function is a polynomial in $z$. The coefficient on the $z^\lambda$ term is precisely $Q_b(N_R, \lambda)$; it is the sum of $\binom{N_R}{\lambda}$ products of the $q_j$'s. $Q_b(N_R, \lambda)$ can be calculated using the residue theorem:
\begin{equation}
	Q_b(N_R, \lambda) = \frac{1}{2 \pi i} \oint{\frac{Z(z)}{z^{\lambda+1}} \ dz}\;,
	\label{eqn:Cm}
\end{equation}
where the contour is around $z=0$.
To estimate this integral, we employ the saddle-point approximation:
\begin{equation}
	Q_b(N_R, \lambda)  \approx \frac{1}{2 \pi i} e^{-f(z_0)} \sqrt{\frac{2 \pi}{f''(z_0)}},
	\nonumber
\end{equation}
where $z_0$ is the (real) value of $z$ at the location of the saddle point of $Z(z)$.  From Eq. \ref{eqn:Cm} it follows that  $f(z)$ is given by $f(z) = (\lambda + 1) \ln{z} - \sum_{j = 1}^{N_R}{\ln{\left(1 + zq_j\right)}}$. The saddle point of $Z(z)$ corresponds to the maximum of $f(z)$ along the real axis. 
The location of this maximum ($z_0$) can be easily computed numerically, allowing one to obtain $Q_b(N_R, \lambda)$. 

This complex residue approach works well for most cases, even when $\lambda$ and $N_R$ are small (provided $N_R$ is greater than approximately $5$). There are, however, a few special cases where the residue approach does not work. However, these are typically the cases where the bound-state partition function can easily be computed exactly:
\begin{enumerate}
\item When $\lambda = 1$ we can immediately write: $Q_b(N_R, \lambda = 1) = \sum_{j = 1}^{N_R}{q_j}$.
\item  When $\lambda = N_R$, we can immediately write $Q_b(N_R, \lambda = N_R) = \prod_{j = 1}^{N_R}{q_j}$.
\item When  $\lambda = N_R - 1$, the function $f(z)$ exhibits no extremum. However, since this corresponds to the case where all receptors but one are bound, it is straightforward to compute $Q_b(N_R, \lambda = N_R - 1) = Q_b(N_R, \lambda = N_R) \times \left(\sum_{j = 1}^{N_R}{\frac{1}{q_j}}\right)$.
\end{enumerate}
When some of the $q_j$ are large, the saddle-point of $f(z)$ moves close to zero and the second derivative becomes extremely large. This can cause numerical errors. Dividing all $q_j$ by the maximum weight $q_{max}$ eliminates this problem. This has the effect of keeping the largest weights in the partition function function near one. The factor $q_{max}^\lambda$ is then re-incorporated into $Q_b(N_R, \lambda)$ afterwards. 

Having found $Q_b(N_R,\lambda)$ it is straightforward to write down the partition function taking the ($N_L-\lambda$) unbound ligands into account. The contribution of the unbound ligands is $Q_{ub}(\lambda) = q_{ub}^{N_L - \lambda}$, where $q_ub$ is the partition function for a single unbound ligand. The complete partition function for $\lambda$ out of $N_L$ ligands binding to $\lambda$ out of $N_R$ receptors, is:
\begin{equation}
	Q(N_R, \lambda) =  \binom{N_L}{\lambda} \lambda! Q_b(N_R,\lambda) Q_{ub}(\lambda).
	\label{eqn:QNR}
\end{equation}
The combinatorial factor in this equation requires an explanation. We consider the situation where $\lambda$ out of $N_L$ ligands bind to $\lambda$ out of $N_R$ receptors. There are $\binom{N_R}{\lambda}$  ways of choosing $\lambda$ out of $N_R$ receptors. Similarly, there are  $\binom{N_L}{\lambda}$ ways to choose $\lambda$ ligands out of $N_L$. However, the total number of distinct ligand-receptor connections is not simply the product of these two binomial factors. The reason is that, for a given choice of $\lambda$ receptors and $\lambda$ ligands, there are still $\lambda !$ ways to make the ligand-receptor connections. We take this into account in the combinatorial factor in Eq. \ref{eqn:QNR}, where we have multiplied  $\binom{N_L}{\lambda}$ by $\lambda !$.

Next, we consider the fact that the $N_R$ receptors can be distributed over the $N_A$ accessible sites. The bound partition function averaged over all distinct ways of connecting $\lambda$ ligands to $\lambda$ surface sites (out of the available $N_A$) is $Q_b(N_A, \lambda) / \binom{N_A}{\lambda}$. This effective partition function is now the binding weight for each of the $\binom{N_R}{\lambda}$ ways of binding $\lambda$ ligands to $N_R$ receptors. Therefore,
\begin{equation}
	\bar{Q}(N_R, \lambda) =  \binom{N_R}{\lambda} \binom{N_L}{\lambda} \lambda! Q_{ub}(\lambda) \frac{Q_b(N_A, \lambda)}{\binom{N_A}{\lambda}}
	\nonumber
\end{equation}
Finally, we average over the probability distribution $P(N_R)$ for observing $N_R$ receptors, and sum over the possible number of bonds $\lambda$ that varies between $0$ and $\min(N_A,N_L)$. This leads to
\begin{align}
	\beta \bar{F} &= -\ln{} \Biggl\{ \sum_{\lambda = 0}^{\min{(N_A,N_L)}} \Biggl[\binom{N_L}{\lambda} \lambda! Q_{ub}(\lambda) \times \nonumber \\
	 &  \frac{Q_b(N_A, \lambda)}{\binom{N_A}{\lambda}} \left(\sum_{N_R = \lambda}^{N_A}{\binom{N_R}{\lambda} P(N_R)}\right)\Biggr] \Biggr\}.
	\label{eqn:Final}
\end{align}

\section{Results}

To test our model, we compute multivalent free energies of binding for two cases. In the first, we consider a nanoparticle coated with mobile ligands, able to bind to a surface with different concentrations of receptors. The second case consists of a particularly simple geometry of $N_L$ ligands grafted to fixed points on a plane, or ``slab'', that can bind to $N_R = N_L$ receptors on the surface. Every ligand has exactly one receptor binding partner, and we can compute the binding free energy analytically. In both cases, we compare to an existing theory of multivalent interactions: the ``valence-limited interaction theory'' (VLIT) \cite{Varilly:2012gl, AngiolettiUberti:2013iu}. A derivation of VLIT for our purposes is given in Appendix A.

\subsection{Multivalent nanoparticle with mobile ligands}

First, we consider a solid ligand-coated nanoparticle adjacent to a surface. The ligands are assumed to be mobile on the particle surface, and chemically identical. The surface, particle core, and ligands are all contained within a three-dimensional simple cubic lattice. 

The ligands are represented as non-self-avoiding lattice walks of $N_{poly}$ steps, beginning anywhere on the particle surface. The weight for a ligand-receptor bond at $j$ is given by $q_j = q'_j e^{-\beta \epsilon}$, where $q'_j$ is the number of walks that terminate at $j$, and the ligand-receptor bond energy $\epsilon$ is set to be the same for all receptors. The quantities $q'_j$ for each surface site are calculated using lattice moment propagation. The partition function $q_{ub}$ per unbound ligand is defined to be the number of non-self-avoiding walks beginning on the particle surface and ending anywhere in the system. Each $q_j$ as well as $q_{ub}$ depend on the particle position $h$ above the receptor surface.

The particle itself is represented as a coarse-grained impenetrable sphere in the lattice, with radius $r$, and center located at $(x^*, y^*, h^*)$. Impenetrability is enforced by preventing any ligand segments from entering lattice sites with coordinates $(x, y, h)$ satisfying $\left(x - x^*\right)^2 + \left(y - y^*\right)^2 + \left(h - h^*\right)^2 \leq r^2$. The receptor surface is located at $h = 1$, and is also impenetrable. 

When the particle is at distance $h$ from the surface, there are $N_A$ sites available to the ligands; this also depends on the choice of ligand length $N_{poly}$. Receptors are placed randomly among the surface sites, and the number $N_R$ placed is chosen from the binomial distribution $P(N_R) = \binom{N_A}{N_R} \phi_R^{N_R} \left(1 - \phi_R\right)^{N_A - N_R}$, where $\phi_R$ is the average receptor concentration.

Our model contains an analytical average over all possible receptor configurations following $P(N_R)$. To calculate the equivalent average free energy of binding in VLIT, we average over many explicit receptor configurations by $\beta \bar{F}_{\text{VLIT}}(h) = -\ln{\left[(1/N) \sum_{n}{e^{-\beta F_n(h)}}\right]}$. Here, $F_n(h)$ is the free energy calculated by VLIT (see Appendix A) for receptor configuration $n$.

\begin{figure}
	\centering
		\includegraphics[width=0.48\textwidth]{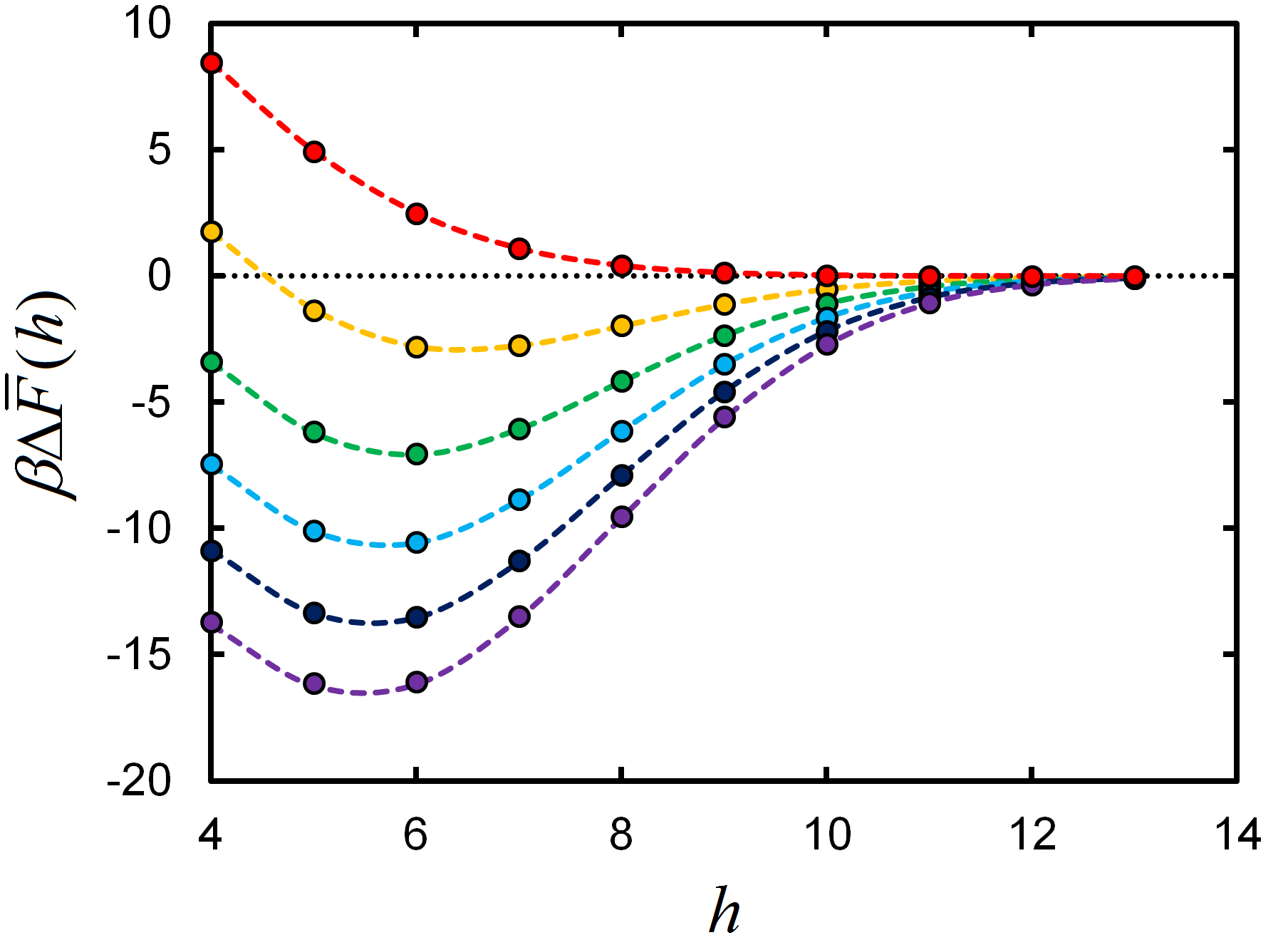}
		\caption{Free energy upon binding $\beta \Delta \bar{F}$ as a function of $h$ for $\phi_R = 0.01, 0.2, 0.4, 0.6, 0.8$, and $1.0$ (red to violet). Points are VLIT results, and lines are fits of our theory, Eq. \ref{eqn:Final}. Fixed parameters are: $N_L = 20$, $\beta \epsilon = -3.5$, particle radius $r = 2$, and $N_{poly} = 20$ segments per ligand. Average number of receptors at $h = 4$ for each dataset, equal to $N_A(h) \phi_R$, is $9.2, 184, 368, 552, 736$, and $920$.}
	\label{fig:FEvsZ}
\end{figure}

We can now directly compare our model results, obtained by Eq. \ref{eqn:Final}, to results from VLIT. This comparison is given in Figure \ref{fig:FEvsZ}. Results are presented as a free energy change $\Delta F(h) = F(h) - F^\circ$ upon surface binding relative to when the particle is infinitely far from the surface. The reference state free energy is given by
\begin{equation}
	\beta F^\circ = - N_L \ln{q^\circ}, 
	\label{eqn:FERefState}
\end{equation}
where $q^\circ$ is the partition function for a single ligand in the reference state.

Results from our model are compared to VLIT averaged over $N = 800$ receptor configurations. For both low and high values of receptor concentration, even when the number of receptors is less than the number of ligands on the particle, our model is in nearly perfect agreement with VLIT. This consistency is encouraging, since for mobile receptors (and/or ligands) VLIT has been proven to become {\it exact} when the number of ligands plus receptors grows large  \cite{stefano-prl}.

\subsection{Multivalent slab with diffuse ligands}

Next, we consider the case where the density of ligands and receptors is so low that one ligand can bind with at most one receptor. This case would correspond to the situation where two surfaces with a low degree of functionalisation interact. In practice, the ligands and receptors would be distributed randomly over the surface. However, to facilitate comparison with analytical theory, we will consider the (trivial) case that each receptor and ligand are directly opposite to each other. 

Hence, we consider a model where $N_L$ ligands are grafted to fixed positions on a slab. The slab is placed at a position $h$ above a surface. A receptor is placed on the surface directly opposite to the tether point of each ligand. The ligands are spaced sufficiently far apart so that each may only bind to the receptor opposite to it. Thus, each ligand is distinct, but has the same binding statistics. As mentioned above: this simplification facilitates comparison with theory, but the numerical method could deal with arbitrary locations of the receptors. This will be discussed shortly.

The ligands are represented by non-self-avoiding walks as in the previous example; however, the partition function $q = q' e^{-\beta \epsilon}$ is now the same for each receptor-bound ligand. The partition function for an unbound ligand, $q_{ub}$, is the number of walks starting at the ligand tether point and ending anywhere in the lattice. 

Since we have assumed that $q$ and $q_{ub}$ are the same for every ligand, we can immediately write an analytical expression for the binding free energy of the slab per ligand: $\beta F(h)^*/N_L = - \ln{\left(q + q_{ub}\right)}$. We now compare the predictions of our model with this analytical result.

Because the number and arrangement of receptors is fixed in this case, we can calculate the binding free energy from our model by simply summing Eq. \ref{eqn:QNR} over all possible values of bound ligands $\lambda$ and taking the natural logarithm:
\begin{equation}
	\beta F(h) = -\ln{\left(\sum_{\lambda = 0}^{N_R}{Q_b(N_R,\lambda) Q_{ub}(\lambda)}\right)}.
	\label{eqn:FEResiduePlane}
\end{equation}
The combinatorial factors $\binom{N_L}{\lambda}$ and $\lambda!$ are not present in this case, as each ligand is distinct. The change in free energy upon binding, $\Delta F(h)$, is calculated relative to the reference state free energy $F^\circ$ when the slab is infinitely far from the surface (Eq. \ref{eqn:FERefState}).

\begin{figure}
	\centering
		\includegraphics[width=0.48\textwidth]{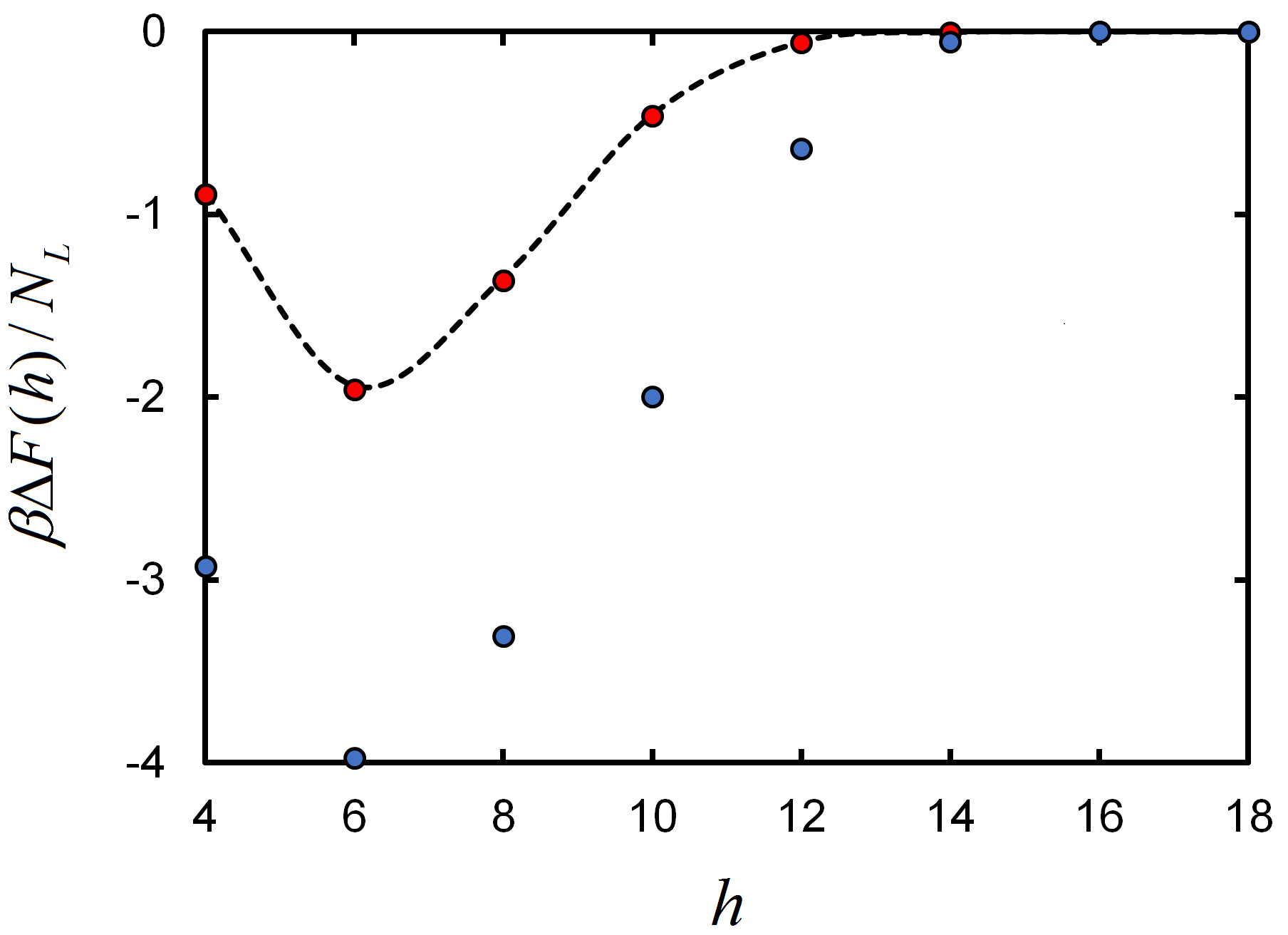}
		\caption{Free energy change per ligand upon slab binding, $\beta \Delta F / N_L$, as a function of slab position $h$. Red points are results from our theory, Eq. \ref{eqn:FEResiduePlane}, and blue points are those from VLIT. The analytical result, $\beta F(h)^*/N_L$, is plotted as the black dashed line. Receptor surface is located at $h = 1$.}
	\label{fig:SlabPlot}
\end{figure}

Figure \ref{fig:SlabPlot} plots values of $\beta \Delta F(h) / N_L$ for several choices of $h$, using $N_L = N_R = 20$, $\beta \epsilon = -7$ and $N_{poly} = 20$ as an example. Results from the VLIT theory (Appendix A) are also given. 

We see that our model agrees almost exactly with the analytical result $\beta F(h)^*/N_L$. Not surprisingly, the VLIT theory does not correctly capture the binding free energy in this particular case, as that theory assumes the probability for any ligand to be unbound is uncorrelated to the probability of a receptor to be unbound. However, when one ligand can only bind one receptor and vice-versa, these probabilities are perfectly correlated, as the two must be bound or unbound at the same time. 

Lack of correlation between unbound ligands and receptors is \emph{not} assumed in our present approach. It therefore applies to the very relevant case where ligands and receptors both have few binding partners.

Importantly, our theory can also be applied to non-trivial cases where each ligand has \emph{different} receptor binding statistics. This may be because each ligand is chemically different, or because they are oriented differently with respect to the surface receptors. The unique receptor binding weight $q_j$ for each ligand is inserted into the auxillary function, Eq. \ref{eqn:AuxFunc}, yielding the bound-state free energy via Eq. \ref{eqn:FEResiduePlane}.

\section{Conclusions}

We have developed a simple approach for computing the binding free energy of a multivalent object on a receptor-coated surface. Using complex residue integration aided by a saddle-point approximation to calculate the bound-state partition function of the multivalent particle yields results that are in nearly exact quantitative agreement with a previously-developed valence-limited interaction theory (VLIT) for multivalent interactions, while also incorporating an analytic average over all local receptor configurations.

Our theory also extends beyond VLIT by providing an accurate multivalent free energy for the case where ligands and receptors have only one binding partner each, even if each ligand-receptor interaction is chemically distinct. This scenario is a specific but non-trivial one, important for a wide range of emerging applications such as interactions between viruses and cell surfaces, as well as nanomedicine \cite{haag}.

The theory presented here holds promise as an easy route to calculating binding free energies in these scenarios, so that large ranges of parameter space may be sampled. It may also serve as a reference point for more detailed simulations of multivalent interactions.

\section{Acknowledgments}
This work has been supported by MC-ITN FP7-People Grant 607602 (``SASSYPOL''), and EPSRC Programme Grant EP/I001352/1.
S. A-U. acknowledges financial support from the Beijing Municipal Government Innovation Center for Soft Matter Science and Engineering. We wish to thank Tine Curk for helpful discussions on this work.

\appendix

\section{Valence-limited interaction theory for a multivalent particle with chemically-indistinct ligands}

The valence-limited interaction theory \cite{Varilly:2012gl, AngiolettiUberti:2013iu} (VLIT) considers two surfaces coated with ligands. The ligands do not interact with each other, with the exception that they may form pairwise bonds with free energy $\beta \Delta f_{jk}$ (where $j$ and $k$ are two ligands). 

The theory does not make any assumptions on the shapes of the two surfaces. Correlations between bonds are introduced by the fact that two ligands cannot bind to a third at the same time.

With these assumptions, the attractive part of the free energy of binding between the two surfaces is given by $\beta F_{att} = \sum_{j}{\left[\ln{p_j} + (1/2)\left(1 - p_j\right)\right]}$, where $p_j$ is the probability that ligand $j$ is \emph{unbound}, and the sum is over all ligands $j$ on the two surfaces \cite{Varilly:2012gl, AngiolettiUberti:2013iu}. 

For each pair of binders $j$ and $k$, $p_j$ must satisfy the relation $p_j + \sum_{k}{p_j p_k \xi_{jk}} = 1$. This is a balance equation, specifying that ligand $j$ must be either unbound (with probability $p_j$), or bound to any of its possible partners $k$ with a free energy of $\beta \Delta f_{jk} = -\ln{\xi_{jk}}$. Note that $p_k$ is the probability that ligand $k$ is unbound.

In our case, all ligands $L$ are identical, while each receptor $j$ is unique, depending on its location on the surface. The balance equations are therefore $p_L + \sum_{j = 1}^{N_R}{p_L p_j \xi_{j}} = 1$ and $p_j + N_L p_j p_L \xi_j = 1$. Here, $p_L$ is the probability that one of the identical ligands is unbound, and $\xi_j$ is the binding weight for receptor $j$, relative to the partition function $q_{ub}$ of a ligand in the unbound state: $\xi_j = q_j / q_{ub}$.

Combining the two relations for $p_L$ yields $p_L + \sum_{j = 1}^{N_R}{(p_L \xi_j)/(1 + N_L p_L \xi_j)} = 1$. This can be solved by a root-finding algorithm. The attraction free energy is then calculated by
\begin{align}
	\nonumber
	&\beta F_{att} = N_L \left[\ln{p_L} + \frac{1}{2} \left(1 - p_L\right)\right] \\
	\nonumber
	& + \sum_{j = 1}^{N_R}{\Biggl\{\frac{1}{2} \left[1 - \left(\frac{1}{1+N_L p_L \xi_j}\right)\right]} - \ln{\left(1 + N_L p_L \xi_j\right)}\Biggr\}
\end{align}
To calculate the total free energy of binding, the repulsive part of the particle/surface interaction must be incorporated. This is equal to the (entropic) repulsion of the ligands due to their proximity to the surface: $\beta F_{rep} = -N_L \ln{q_{ub}}$. The total binding free energy is then $F = F_{att} + F_{rep}$.

\bibliography{main}

\

\

\

\

\

\

\

\end{document}